%% file: main.tex
\documentclass[reprint,
superscriptaddress,frontmatterverbose,showpacs,preprintnumbers,showkeys,nofootinbib,
nobibnotes,amsmath,amssymb,pra,floatfix,twocolumn
]{revtex4-1}
\usepackage{graphics}
\usepackage{ae}
\usepackage[T1]{fontenc} 
\usepackage[english]{babel}
\usepackage[utf8]{inputenc}
\usepackage{amsmath}
\usepackage{graphicx}
\usepackage{color}
\usepackage{dcolumn}
\usepackage{bm}
\usepackage{hyperref}
\usepackage[mathlines]{lineno}
\usepackage{xspace}
\usepackage{algorithm}
\usepackage{algorithmic}
\usepackage{subfigure}
\usepackage{amsfonts}
\usepackage{amsmath}
\usepackage{enumerate}
\usepackage{framed}
\usepackage{bm}
\usepackage{amsthm}
\input{epsf}

\renewcommand{\eqref}[1]{{Eq.~(\ref{#1})}}

\usepackage[utf8]{inputenc}
\usepackage{amsmath}
\usepackage{amsthm}
\usepackage{eucal}
\usepackage{amssymb}
\usepackage{mathrsfs}
\usepackage{physics}
\usepackage{graphicx} 
\begin{document}
\title{Ivancevic Option Pricing Model modulational instability through the variational approach}

\author{Christopher Gaafele}
	\email {christopher.gaafele@studentmail.biust.ac.bw}
	\affiliation{Department of Physics and Astronomy, Botswana International University of Science and Technology, Private Mail Bag 16 Palapye, Botswana}
\date{June 2024}

\begin{abstract}
    The instability of the Ivancevic option pricing model is studied through the variational method. We have analytically derived the dispersion relation of the IOPM for both constant volatility and Landau coefficient model and time-dependent volatility and Landau coefficient model. Also the IOPM was studies numerically using the 4th order Runge-Kutta method.
\end{abstract}

\maketitle
\section{Introduction}
\label{sec1}
Since the discovery of the Nonlinear Schrodinger equation (NLS) like option pricing model in 2010, it has drawn much interest theoretically in various soliton phenomena in the financial world. Variuos techniques such as the generalized Kudryashov method (GKM)~\cite{demiray2022new}, adomian decomposition method (ADM) and spectral boundry approximation (SBA) method~\cite{adama2022exact}, improved physics-informed neural network (IPINN)~\cite{bai2022application}, the method of Lie symmetries~\cite{obaidullah2023analytical}, the Elsaki-Adomian decomposition method~\cite{gonz2015solving}, projected differential transform method (PDTM)~\cite{edeki2017analytical}, He's frecuency amplitude formulation~\cite{gonzalez2017solving}, \(q-homotopy analysis transform method\) (\(q\)-HATM)~\cite{veeresha2022analysis} and modulational instability (MI) analysis ~\cite{chen2022soliton}~\cite{gaafele2024modulational} among others~\cite{raheel2023research}~\cite{wroblewski2023quantum}.

Variational approach (VA) has also been a  versatile method for nonlinear like models for different branches of physics including optics~\cite{anderson1983variational}~\cite{malomed1997pulse}~\cite{malomed2002variational} and in Bose-Einstein condesates~\cite{sabari2013variational}~\cite{wamba2008variational}~\cite{wamba2014dynamical}. In most studies this analysis has the same results as MI and other method that are used for the NLS like models, in terms of soliton dynamics and stability. Unlike MI, VA is suitable in discribing soliton pulse propagation. VA can either use \(Gaussian\) ansatz~\cite{nicolin2008nonlinear} or \(sech\) ansatz~\cite{malomed2002variational} among others since they are localize and assumes the shape of a soliton. In this project we consider modulated wave ansatz, which is the same wave ansatz used in modulational instability as shown by~\cite{wamba2008variational}~\cite{sabari2013variational}~\cite{wamba2014dynamical}~\cite{ndzana2007modulational}. The same can be done for the IOPM, and several quantities such as the effective Hamiltonian and effective potantial can be derived for the stock/asset market system, which is as important as MI in the financial world~\cite{chen2022soliton}~\cite{gaafele2024modulational}.   

The IOPM is the NLS like model used in asset derivative markets alternative approach to the traditional Black-Scholes (BS) model for pricing options. Developed by Vladimir Ivancevic, this model incorporates elements of both classical mechanics and financial mathematics~\cite{ivancevic2010adaptive}.  It uses a Hamiltonian formalism to describe the dynamics of the underlying asset price and employs a different stochastic process than the standard geometric Brownian motion used in the Black-Scholes model. The model incorporates nonlinear dynamics, which can better capture the real-world behavior of financial markets, including features such as volatility clustering and heavy tails in the distribution of returns. Ivancevic’s approach sometimes draws analogies from quantum mechanics, using path integrals to represent the probabilities of different price paths~\cite{kleinert2009path}. This can provide a more flexible and accurate representation of price movements over time. The Black-Scholes equation is a linear partial differential equation, while the Ivancevic model often results in nonlinear equations due to its more complex stochastic processes. The IOPM can easily incorporate quantum neural computations~\cite{ivancevic2009quantum}. 

The remaining section are as follows: Section \ref{sec2} will focus on the Ivancevic option pricing model. Section \ref{sec3} will focus on the variational calculation of the model. Section \ref{sec4} shows direct simulation of the ansatz for different conditions and the paper is concluded by Section \ref{sec5}.

\maketitle
\section{The Model}
\label{sec2}

The classic Black-Scholes (BS) model is a significant breakthrough in financial mathematics, capturing the dynamic behavior of market prices for financial assets, such as stock options, over time~\cite{black1973pricing}. This model operates under certain assumptions, where variables such as the asset price\(s\), drift parameter \(\mu\), and volatility \(\sigma\) are considered constants. Additionally, it presumes the existence of frictionless, arbitrage-free, and efficient markets~\cite{gonzalez2016nonlinear},\cite{edeki2016modified}. However, to address the limitations of these assumptions, various modifications have been suggested, including stochastic interest rate models, jump-diffusion models, stochastic volatility models, and models incorporating transaction costs. As a result, the Ivancevic option pricing model (IOPM) has gained increasing attention in recent years for its effectiveness in estimating option values, serving as an alternative to the Black-Scholes equation\cite{chen2022soliton},\cite{ali2023physical},\cite{elmandouh2022integrability},\cite{chen2021dark},\cite{gonzalez2017solving}. The price function \(s = s(t)\) for \(0 \leq t \leq T\) and follows the geometric brownian motion (BM) \(ds = s(\mu dt + \sigma dW_t)\), where \(\mu\) is the drift rate, \(W_t\equiv W(t)\) is the Weiner process. The BS model is given by

\begin{equation}
    \frac{\partial\psi}{\partial t} + rs\frac{\partial\psi}{\partial s} + \frac{1}{2}\sigma^2s^2\frac{\partial^2\psi}{\partial s^2} - r\psi = 0.
\label{Eq1}
\end{equation}
Although there are versions of the Black-Scholes (BS) model tailored for American stock regulations, the aforementioned BS model applies to European stock regulations, where the option is exercised at maturity \(T\). Recently, Vukovic~\cite{vukovic2015interconnectedness} established a connection between the Schrödinger equation (SE) and the Black-Scholes model equation using principles from quantum physics, particularly the Hamiltonian operator. It was noted that the BS equation could be derived from the SE using tools from quantum mechanics~\cite{contreras2010quantum}. While the SE is a complex-valued equation, the BS model is a real-valued equation governing the price function. The BS model equation, Eq.(\ref{Eq1}), can be extended to one-dimensional option models involving \(\psi\) and \(s\). As discussed in~\cite{voit2003statistical}, one can determine the probability density function (PDF) by solving the Fokker–Planck equation using classical Kolmogorov probability methods, rather than relying on the option value obtained from the BS equation.

In the contemporary era, exploring the intricate aspects of economic and financial issues through mathematical modeling has become a highly researched domain due to its extensive applications in nonlinear science. These mathematical models often take the form of nonlinear partial differential equations (NPDEs). Among the widely studied NPDEs is the Ivancevic option pricing model. Ivancevic~\cite{ivancevic2010adaptive} employed quantum-probabilistic principles to derive a probability density function (PDF) equivalent for the value of a stock option by constructing a complex-valued function. This approach resulted in the proposal of a nonlinear model~\cite{cont2001empirical}, which was subsequently renamed the Ivancevic Option Pricing Model (IOPM), as outlined below

\begin{equation}
    i\frac{\partial\psi}{\partial t} + \frac{1}{2}\sigma^2\frac{\partial^2\psi}{\partial s^2} + \beta|\psi|^2\psi = 0.
\label{Eq2}
\end{equation}
In this context, \(\psi(s,t)\) represents the option price function at time \(t\), and \(\sigma\) is the volatility, which remains constant in this model. The adaptive market potential, represented by \(\beta\), is the Landau coefficient indicating the adaptive market potential. In its simplest form, it is equivalent to the interest rate \(r\); however, in this case, it depends on a set of adjustable parameters \(\{{w_i}\}\), known as the synaptic weight vector components. This adaptive market potential is related to market temperature, which follows the Maxwell-Boltzmann distribution~\cite{kleinert2009path}. Additional relevant studies have been referenced to enhance understanding of the current analysis. Edeki et al.~\cite{edeki2017analytical} and Gonzalez-Gaxiola and Ruiz de Chavez~\cite{gonz2015solving} investigated the Ivancevic option pricing model using analytical approaches, specifically the projected differential transformation method and a hybrid technique combining the Adomian decomposition method with the Elsaki transform, respectively. Chen et al.~\cite{chen2019new} introduced a novel operator splitting method to solve the fractional Black-Scholes model for American options.

\maketitle
\section{Variational approximation}
\label{sec3}
We perform a variational approach on the IOPM given by Eq.(\ref{Eq2}). The lagrangian density of the IOPM is given by

\begin{equation}
    \mathcal{L} = \frac{i}{2}\left(\psi^*\frac{\partial\psi}{\partial t} - \psi\frac{\partial\psi^*}{\psi}\right) - \frac{1}{2}\sigma^2\left|\frac{\partial\psi}{\partial s}\right|^2 + \frac{1}{2}\beta|\psi|^4.
\label{Eq3}
\end{equation}

we assume the modulational instability motivated wave ansatz given by 

\begin{equation}
    \psi(s,t) = [\psi_0 + a_1(t)e^{i(b_1(t) + qx)} + a_2(t)e^{-i(b_2(t) - qx)}]e^{i(kx - \omega t)},
\label{Eq4}
\end{equation}

we substitute the ansatz into the lagrangian density to get

\begin{multline}
    L = \pi[-2(a_1^2\frac{db_1}{dt} + a_2^2\frac{db_2}{dt})\\ + 2(\frac{1}{2}\sigma^2k^2-\beta\psi_0^2)(\psi_0^2+a_1^2+a_2^2)\\ + \beta(\psi_0^4+a_1^4+a_2^4+4\psi_0^2(a_1^2+a_2^2)+4a_1^2a_2^2+4a_1a_2\psi_0^2cos(b_1+b_2))\\ - \sigma^2(k^2\psi_0^2 + k^2(a_1^2+a_2^2) + (q^2+2kq)(a_1^2-a_2^2))],
\label{Eq5}
\end{multline}

Where \(L = \int_{0}^{2\pi}\mathcal{L}ds\). From Eq.(\ref{Eq5}) \(b_1\),\(b_2\) are generalized coordinates of the system and \(2a_1^2\), \(2a_2^2\) are generalized momenta and are canonically conjugate with respect to the effective Hamiltonian where \(A_1(t) = 2a_1^2(t)\) and \(A_2(t) = 2a_2^2(t)\). The effective Hamiltonian is given by 

\begin{multline}
    H_{eff} = (\beta\psi_0^2 - \frac{1}{2}\sigma^2q^2)(A_1 + A_2) - kq\sigma^2(A_1 - A_2)\\ + 2\beta\psi_0^2\sqrt{A_1A_2}cos(b_1 + b_2) + \frac{1}{4}\beta(A_1^2 + A_2^2 + 4A_1A_2).
\label{Eq6}
\end{multline}

We use the Euler-Lagrange equation of motion based on the Lagrangian Eq.(\ref{Eq5}), which are given by

\begin{equation}
    \frac{d}{dt}\frac{\partial L}{\partial\dot{\xi}_{i}} - \frac{\partial L}{\partial\xi_{i}} = 0
\label{Eq7}
\end{equation}
where \(\xi_{i}\) is the generalized coordinates. The evolution of the variational equations \(a_1, b_1, a_2, b_2\) are given by

\begin{align}
    \frac{db_1}{dt} &= \beta\psi_0^2 - \frac{1}{2}\sigma^2q(q+2k), \label{Eq8}\\
    \frac{db_2}{dt} &= \beta\psi_0^2 - \frac{1}{2}\sigma^2q(q+2k), \label{Eq9}\\
    \frac{da_1}{dt} &= a_2\psi_0^2sin(b_1 + b_2), \label{Eq10}\\
    \frac{da_2}{dt} &= a_1\psi_0^2sin(b_1 + b_2). \label{Eq11}
\end{align}
Assuming that \(b = b_1 + b_2\), we get
\begin{align}
    a_1 &= a_2, \label{Eq12}\\
    \frac{da_1}{dt} &= \beta\psi^2a_1sin(b), \label{Eq13}\\
    \frac{db}{dt} &= 2\beta\psi_0^2 - 2q^2 + 2\beta\psi_0^2cos(b).
\label{Eq14}
\end{align}

The solution to Eq.(\ref{Eq14}) is given by 
\begin{widetext}
\begin{equation}
    b = 2arctan\left[\frac{\sqrt{(2\beta\psi_0^2 - \frac{1}{2}q^2\sigma^2)(-\frac{1}{2}q^2\sigma^2)}}{-q^2P}tan\sqrt{(2\beta\psi_0^2 - \frac{1}{2}q^2\sigma^2)(-\frac{1}{2}q^2\sigma^2)t}\right].
\label{Eq15}
\end{equation}
\end{widetext}

One can notice that Eq.(\ref{Eq15}) is used to solve Eq.(\ref{Eq13}). And also if \(q^2 - 4\frac{\psi_0^2}{\sigma^2}\beta<0\) the solution of Eq.(\ref{Eq13}) is given by

\begin{equation}
    a = a_0\sqrt{1 - 4\frac{\beta\psi_0^2}{q^2\sigma^2}sin^2\left(\sqrt{\left(\frac{1}{2}q^2\sigma^2 - 2\beta\psi_0^2\right)\left(-\frac{1}{2}q^2\sigma^2\right)}t\right)},
\label{Eq16}
\end{equation}
while if \(q^2 - 4\frac{\psi_0^2}{\sigma^2}\beta>0\) we get 
\begin{equation}
    a = a_0\sqrt{1 + 4\frac{\beta\psi_0^2}{q^2\sigma^2}sinh^2\left(\sqrt{\left(2\beta\psi_0^2 - \frac{1}{2}q^2\sigma^2\right)\left(-\frac{1}{2}q^2\sigma^2\right)}t\right)}.
\label{Eq17}
\end{equation}
 The \(2\) equations given by Eq.(\ref{Eq16}) and Eq.(\ref{Eq17}) are sinusoidal and exponential respectively signifying stability and instability of the model. 

\begin{figure}[h]
    \centering
    \begin{minipage}{0.49\textwidth}
        \includegraphics[width=1\textwidth]{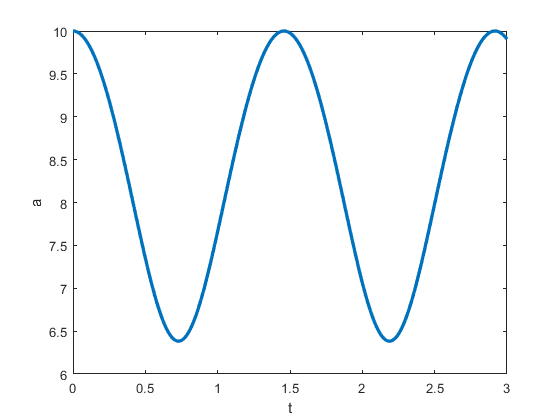}
        \caption{The panel shows the time evolution of \(a\) as a function of \(t\) where \(\beta = 1\), \(\sigma = 1.5\), \(q = \sqrt{3}\), \(\psi_0 = 1\) for Eq.(\ref{Eq16})}
        \label{fig:2.0}
    \end{minipage}
    \hfill
    \begin{minipage}{0.49\textwidth}
        \includegraphics[width=1\textwidth]{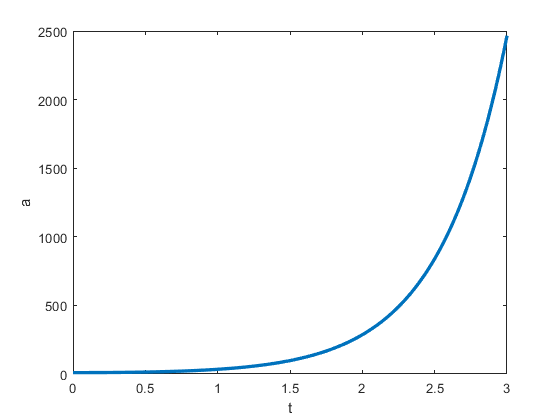}
        \caption{The panel shows the time evolution of \(a\) as a function of \(t\) where \(\beta = 1\), \(\sigma = 1.5\), \(q = \sqrt{3}\), \(\psi_0 = 1\) for Eq.(\ref{Eq16})}
        \label{fig:2.1}
    \end{minipage}
    \label{fig:19972019}
\end{figure}

We can reduce Eqs.(\ref{Eq8})-(\ref{Eq11}) to the realm of one degree of freedom with an effective potential energy landscape whose variation gives a more vivid understanding of the instability. We also assume that \(A_1(t = 0) = 0\) in Eq.(\ref{Eq5}) and \(A_1(t) = A_2(t)\), we have

\begin{equation}
    \dot{H} = \pi\left[2\left(\beta\psi_0^2 - \frac{1}{2}\sigma^2q^2\right) + 2\beta\psi_0^2Acos(b) + \frac{3}{2}\beta A^2\right].
\label{Eq18}
\end{equation}
Eliminating \(b\) for \(A(t=0)=0\) we get the mechanical energy equation for \(A\)
\begin{equation}
    \frac{\dot{A}^2}{2} + V_{eff} = 0,
\label{Eq19}
\end{equation}
where the effective potential \(V_{eff}(A)\) is given by
\begin{equation}
    V_{eff} = -2A^2\left(\beta\psi_0^2\right)^2 - \left(2\left(\beta\psi_0^2 - \frac{1}{2}\sigma^2q^2\right) + \frac{3}{2}\beta A\right)^2.
\label{Eq20}
\end{equation}
We can examine the stability of the effective potential by evaluating its curvature at \(A=0\), we get \(V_{eff}^" = 2\sigma^2q^2\left(\frac{1}{2}\sigma^2q^2 - 2\beta\psi_0^2\right)\), which is convex for \(\sigma^2q^2>4\beta\psi_0^2\), thus describing stable dynamics, and \(V_{eff}\) is concave for \(\sigma^2q^2<4\beta\psi_0^2\), thus describing unstable dynamics. 

\begin{figure}
    \centering
    \includegraphics[width=0.49\textwidth]{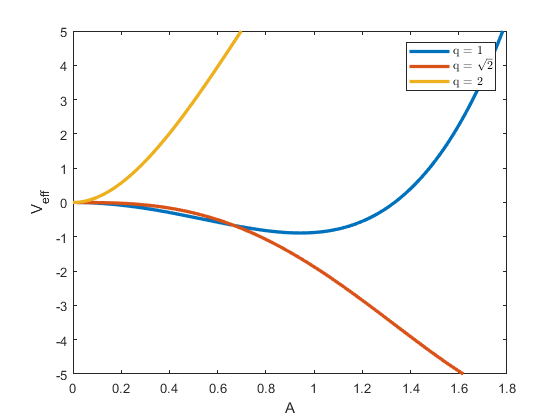}
    \caption{The panel shows the effective potentential given by Eq.(\ref{Eq20}) for different values of \(q\)}
    \label{fig:3.0}
\end{figure}

\subsection{Variational Approach for time depended volatility and adaptive market heat potential}

In this section we study the variational approach for IOPM Eq.(\ref{Eq2}) with time dependent volatility \(\sigma\equiv\sigma(t)\) and the adaptive market heat potential \(\beta\equiv\beta(t)\), given by

\begin{equation}
    i\frac{\partial\psi}{\partial t} + \frac{1}{2}\sigma(t)^2\frac{\partial^2\psi}{\partial s^2} + \beta(t)|\psi|^2\psi = 0.
\label{Eq21}
\end{equation}

This approach is important since it assumes the volatility is non-constant which is the limitation of the BS and IOP models, since volatility is not constant in the real world. Also in this case the adaptive market heat potential is assumed to be non-constant.

We consider 

\begin{equation}
    \psi = \psi_1[1 + \eta(t)cos(qs)],
\label{Eq22}
\end{equation}
where \(\psi_1 = e^{i(-k^2\int_{0}^{t}-\frac{1}{2}\sigma(t')^2dt' + \int_{0}^{t}\beta(t')dt' + ks)}\), without loss of generality, we assume that \(a_1 = a_2\), \(b_1 = b_2\) and \(\psi_0 = 1\). Following the same procedure that we used above, we get 

\begin{align}
    \dot{\eta}_r &= \frac{1}{2}q(t)^2\sigma^2\eta_i - \frac{3}{4}\beta(t)(\eta_i^3 + \eta_i\eta_r^2),\label{Eq23}\\
    \dot{\eta}_i &= -(\frac{1}{2}q(t)^2\sigma^2 - 2\beta(t))\eta_r + \frac{3}{4}\beta(t)(\eta_r^3 + \eta_r\eta_i^2),\label{Eq24}
\end{align}
for a stability equation \(w(t) = w_r(t) + iw_i(t)\), hence, at the linear level, we get 

\begin{equation}
    \ddot{\eta} = \frac{\dot{\sigma}(t)}{\sigma(t)}\dot{\eta}_r - \frac{1}{2}\sigma(t)^2q^2\left[\frac{1}{2}\sigma(t)^2q^2 - 2\beta(t)\right]\eta_r.
\label{Eq25}
\end{equation}
We can determine the modulational stability by determining the criterion of stability of the ordinary differential equation(\ref{Eq25}). In the case of constant \(\sigma(t)\) and periodic \(\beta(t)\) Eq.(\ref{Eq25}) becomes Hill's equation, which is well studied~\cite{ndzana2007modulational}. Eq.(\ref{Eq25}) becomes the Mathieu equation for \(\beta(t) = 2\alpha cos(2\omega t)\) where \(\alpha = \beta\psi_0^2 - \frac{1}{2}\sigma^2q^2 - \sigma^2qk\) which is also well studied. In this case Eq.(\ref{Eq25}) becomes

\begin{equation}
    \ddot{\eta} + \frac{1}{2}\sigma^2q^2[\frac{1}{2}\sigma^2q^2 - 4\alpha cos(2\omega t)]\eta_r = 0,
\label{Eq26}
\end{equation}
after some approximation Eq.(\ref{Eq27}) becomes

\begin{equation}
    \ddot{\eta} = Q^2[1 + \alpha_1'cos(2\omega t)]\eta_r = 0,
\label{Eq27}
\end{equation}
where \(Q^2 = \frac{1}{4}\sigma^4q^4\), \(\alpha_1'= -\frac{8\alpha_1}{\sigma^2q^2}\). The non perturbed version of Eq.(\ref{Eq27}) is given by

\begin{equation}
    \ddot{\eta} + Q^2\eta_r = 0.
\label{Eq28}
\end{equation}
The solurion of Eq.(\ref{Eq28}) is given by

\begin{equation}
    \eta_r = \eta_0e^{iQt} + c.c,
\label{Eq29}
\end{equation}
where \(\frac{Q}{2\pi}\) is the fundamental frequency of the system. Therefore the perturbed solution of Eq.(\ref{Eq27}) is given by

\begin{equation}
    \eta(t) = e^{(\mu + i\omega)t + ib} + e^{(\mu - i\omega)t - ib},
\label{Eq30}
\end{equation}
of which when substituted into Eq.(\ref{Eq27}) gives us

\begin{equation}
    \frac{1}{2}Q^2\alpha_1' e^{-ib} + [(\mu + i\omega)^2 + Q^2] e^{ib} = 0.
\label{Eq31}
\end{equation}

The solution of the above equation  is such that

\begin{equation}
    X^2 + 2Q^2X + C = 0,
\label{Eq32}
\end{equation}
where \(X = \mu^2 + \omega^2\) and \(C = -4\omega^2Q^2 + \frac{Q^4}{4}(4-\alpha_1'^2)\). This quadratic equation, yields the solutions 
\begin{align}
    X_1 &= -Q^2-2\omega Q - \frac{Q^3}{16\omega}\alpha_1'^2,\label{Eq33}\\
    X_2 &= -Q^2-2\omega Q+\frac{Q^3}{16\omega}\alpha_1'^2.\label{Eq34}
\end{align}

We can use the above equation to get 
\begin{equation}
    \mu^2 = -Q^2 - \omega^2 + 2\omega Q + \frac{Q^3}{16\omega}\alpha_1'^2.
\label{Eq35}    
\end{equation}
And \(Q = \frac{1}{2}m\Gamma\) for \(m = 1,2,3,\dots\) are defined as the parametric resonances, where \(\Gamma = 2\omega\). We get 
\begin{equation}
    Q^2 + \omega^2 - 2\omega Q = \left(\frac{1}{2}\Gamma + \epsilon_1\right)^2 + \frac{\Gamma}{4} - \Gamma\left(\frac{1}{2}\Gamma + \epsilon_1\right) = \epsilon_1^2,
\label{Eq36}
\end{equation}
by assuming \(U = -Q^2 - \omega^2 + 2\omega Q\). For \(Q = \frac{1}{2}\Gamma + \epsilon_1\) we get 

\begin{equation}
    \lambda_1^2 = \frac{1}{16}Q^2\alpha_1'^2 - \epsilon_1^2,
\label{Eq37}
\end{equation}

for \(m = 1\), \(\mu = \lambda_1\) for Eq.(\ref{Eq36}) and Eq.(\ref{Eq37}) is called the value of the instability increment for the first resonance. We can clearly observe that the point of maximum instability coincides with the width of the resonance

\begin{align}
    \lambda_{1max}^2 &= \frac{1}{64}\sigma^4q^4\alpha_1'^2,\label{Eq38}\\
    \epsilon_1^2 &= \frac{1}{64}\sigma^4q^4\alpha_1'^2,\label{Eq39}.
\end{align}

The second parametric resonance appears for m =2, 

\begin{equation}
    Q = \Gamma + \epsilon_2.
\label{Eq40}
\end{equation}
We get the increment of instability \(\lambda_2\) for the second domain is given by

\begin{equation}
    (4\lambda_2)^2 = \left(\frac{\alpha_1'^2}{2}\right) + \alpha_2'^2 - \left(\frac{\alpha_1^2}{3} - 4\epsilon_2\right)^2,
\label{Eq41}
\end{equation}
which is maximum at \(\epsilon_2 = 0\). The edges of this instability region are determined by the condition \( \lambda_2 = 0 \) as

\begin{align}
    \epsilon_2^+ &= \frac{1}{24}\alpha_1'^2,\label{Eq42}\\
    \epsilon_2^- &= \frac{-5}{24}\alpha_1'^2.\label{Eq43}
\end{align}

It is clear from Eq.(\ref{Eq42}) and Eq.(\ref{Eq43}) that the instability region is not symmetrical around the center frequency.

It can be concluded that a slight periodic variation in the volatility of the stock price generates new regions of modulation instability.

\maketitle
\section{Numerical Simulations}
\label{sec4}
In this section we study the IOPM Eq.(\ref{Eq2}) by direct numerical algoritms. We use the improved Euler's method refered to as the 4th order Runge-Kutta method. In this method time is discrete and increases in time steps \(\Delta t = \frac{1}{4000}\). And also the initial solution is given by \(\psi(s,0) = \sqrt{n} + \epsilon cos(ks)\). In the algorithm we use \(n = 10\), \(\epsilon = 0.01\) and \(k=0.5\). 

\begin{figure}[h]
    \centering
    \begin{minipage}{0.35\textwidth}
        \includegraphics[width=1\textwidth]{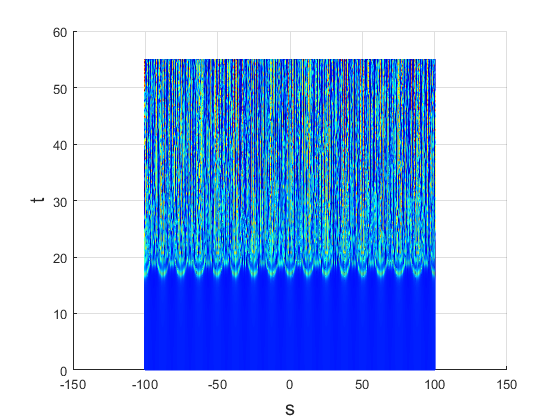}
        \caption{The Figure shows the temporal development of the CW wave of the option price $\psi$ for \(\sigma=0.2\), \(\beta=2\). }
        \label{fig:4.1}
    \end{minipage}
\end{figure}
\begin{figure}[h]
    \begin{minipage}{0.35\textwidth}
        \includegraphics[width=1\textwidth]{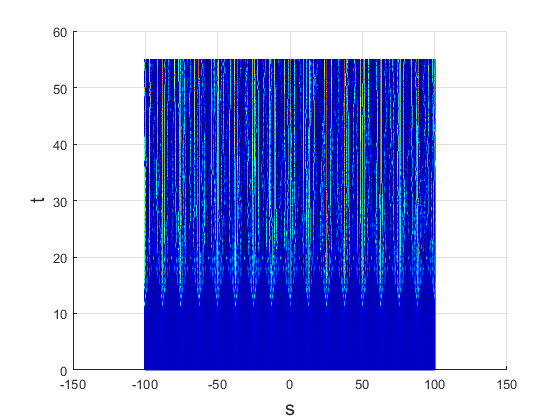}
        \caption{The Figure shows the temporal development of the CW wave of option price $\psi$ for \(\sigma=0.4\), \(\beta=2\).}
        \label{fig:4.2}
    \end{minipage}
\end{figure}
\begin{figure}[h]
    \begin{minipage}{0.35\textwidth}
        \includegraphics[width=1\textwidth]{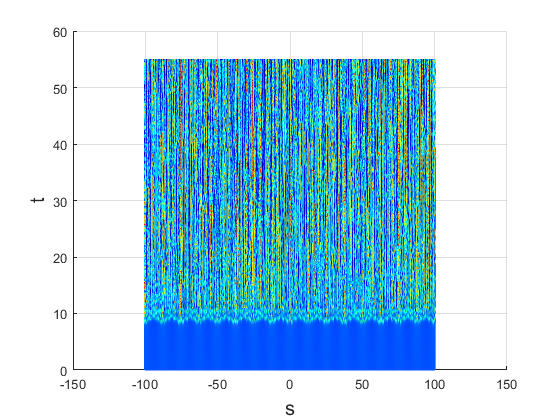}
        \caption{The Figure shows the temporal development of the CW wave of option price $\psi$ for \(\sigma=0.2\), \(\beta=4\).}
        \label{fig:4.3}
    \end{minipage}
\end{figure}
\begin{figure}[h]
    \begin{minipage}{0.35\textwidth}
        \includegraphics[width=1\textwidth]{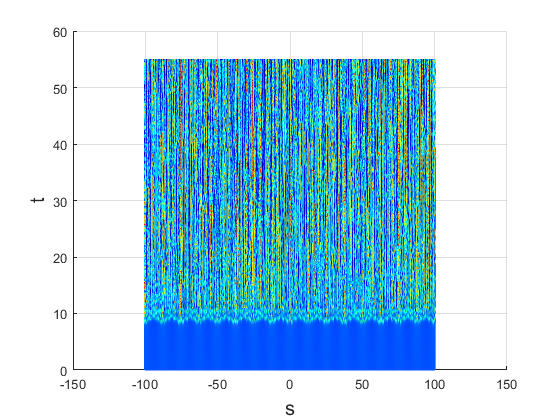}
        \caption{The Figure shows the temporal development of the CW wave of option price $\psi$ for \(\sigma=0.4\), \(\beta=4\).}
        \label{fig:4.4}
    \end{minipage}
\end{figure}

The direct numerical simulations as shown by figure(\ref{fig:4.1})-(\ref{fig:4.4}) shows that an increase in both the volatility and the adaptive market heat potential leads to the instability of the CW wave starting sooner as we can see on figure(\ref{fig:4.1}) that the solitons begin to emerge at time close to \(20\) when \(\sigma = 0.2\) and \(\beta = 2\) which is smaller compared to on figure(\ref{fig:4.4}) where \(\sigma = 0.4\) and \(\beta = 4\), the solitons emerge at \(t = 10\). Which agrees with MI results from optics~\cite{dauxois2006physics}.

\maketitle
\section{Conclusion}
\label{sec5}
We have studied the modulational instability of the IOPM using the variational method. this method also derives a dispersion relation, also this same method also studies the stability criterion of the IOPM with non-constant volatility and the adaptive market heat potential, which gives new results different from other studies given by previous results on the IOPM whose concern's were of constant volatility and adaptive market heat potential. Also we have used direct numerical analysis on the IOPM.

\include{bib}
\end{document}